\documentclass[preprint2]{aastex}

\begin{document}

\title{The usefulness of 2MASS JHK$_{\rm s}$ photometry \\ for open cluster studies}
\author{David G. Turner}
\affil{Saint Mary's University, Halifax, Nova Scotia, Canada}
\email{turner@ap.smu.ca}

\begin{abstract}
2MASS {\it JHK}$_s$ data are used to infer the reddening and distance of open clusters for which limited optical data are available. Intrinsic ZAMS color-color and color-magnitude relations are derived with reference to existing calibrations, standard stars, three uniformly-reddened clusters: Stock 16, NGC 2362, and NGC 2281, and unreddened Hyades dwarfs. The method of inferring interstellar reddening and distance for sparsely-populated open clusters is applied to Berkeley 44, Turner 1, and Collinder 419, for which existing results conflict with those inferred from {\it JHK}$_s$ data. The last two clusters are of special interest: Turner 1 because it hosts the Galaxy's longest-period classical Cepheid, and Collinder 419 because it lies in the Cygnus X complex.
\end{abstract} 

\section{Introduction}
The introduction of CCD detectors to photometric studies of open clusters has led to significant improvements in the accuracy and precision of data for faint cluster stars observed previously via photoelectric or photographic techniques. The tradeoff is a low efficiency and associated steep wavelength dependence for such detectors in the blue-violet region, limiting photometric accuracy in the traditional ultraviolet filters: the Johnson system {\it U}-band and the Str\"{o}mgren system {\it u}-band. Color corrections for observations in the ultraviolet become non-linear and multi-valued in that situation, particularly for hot stars that display a sizable Balmer discontinuity superposed on their continuum \citep[see, for example,][]{mv77}. Many recent open cluster studies have therefore been restricted to observations in the {\it BVRI} bands, or simply the {\it BVI} or {\it VRI} bands.

The intrinsic relations for OB stars and GK dwarfs in color-color diagrams restricted to such systems are nearly parallel to the reddening vectors for interstellar extinction \citep{ca93}, posing difficulties for establishing the reddening of cluster stars \citep[see][]{ca09}. AF stars can be studied in such fashion \citep[e.g.,][]{te11}, but that would limit photometric studies in {\it BVRI} to intermediate-age clusters with their main sequences consisting of unevolved 1--2 $M_{\odot}$ stars. Even in such cases, an accurate knowledge of the interstellar reddening of member stars is essential for the reliable establishment of distance and age \citep{tu96}, which means that the photometric study of many clusters may be limited by the nature of the filter bands employed for the observations. Some researchers studying intermediate-age and old clusters have therefore adopted an alternate approach to establish cluster intrinsic parameters, by identifying a putative red giant clump in cluster color-magnitude diagrams arising from the He-burning stage of low-mass red giants \citep{ca70} and inferring the reddening, age, and distance of a cluster through optimized fitting of model isochrones to the data \citep*[e.g.,][]{ca95,ca04}.

There are two problems associated with such an approach. First, the dependence of the fitting technique on model isochrones may bias the results. Second, the most frequently encountered stars along most Galactic lines of sight tend to be K giants and A dwarfs \citep{mc65,mc70}, which are also the most luminous members of intermediate-age open clusters. As noted by \citet{bm73}, simulations of {\it UBV} photometry for unrelated stars lying in typical Galactic star fields can generate color-magnitude diagrams containing pseudo-main sequences similar to those of true open clusters. The presence of physically-unrelated A dwarfs and K giants along typical Galactic lines of sight may therefore generate features in open cluster color-magnitude diagrams, A dwarfs and G giants or FG dwarfs and K giants, in similar locations to those found in the color-magnitude diagrams of intermediate-age or old open clusters, respectively. The photometric properties of foreground or background stars in some open cluster fields could therefore be mistaken for the characteristics of an old open cluster color-magnitude diagram, thereby biasing studies of clusters tied to the red clump method.

An independent means of establishing cluster membership (e.g., star counts, proper motions, or radial velocities) and especially reddening for cluster stars is therefore essential for avoiding incorrect conclusions about cluster parameters. The ready availability of {\it JHK}$_s$ photometry \citep{cu03} from the Two Micron All Sky Survey \citep[2MASS,][]{sk06} for Galactic star fields offers one such means, since it addresses the primary parameter vital for open cluster studies: the amount of foreground reddening of cluster stars \citep*[e.g.,][]{ma08}. Presented here are examples of star clusters, Berkeley 44, Turner 1, and Collinder 419, for which the method addresses potential incorrect choices of cluster parameters based solely on optical photometry.

\section{{\it JHK}$_{s}$ Intrinsic Relations}
The effective wavelengths for the 2MASS {\it JHK}$_s$ system, 1.235 $\mu$m, 1.662 $\mu$m, and 2.159 $\mu$m, respectively, are fairly close to those for the older {\it JK} system of \citet{jo68} and the {\it JHK} system studied by \citet{ko83}, and observations with the {\it JHK} and {\it K}$_s$ filters are standardized in fairly similar fashion. The main source of difficulty is likely to be the presence of atmospheric absoprtion features within the same photometric bands \citep{my05,my08}, which reduces the precision of repeated observations. {\it K}$_s$-band observations, and possibly {\it H}-band observations, of stars are also suceptible to emission from circumstellar dust. Is it possible that the intrinsic {\it VJK} and {\it VJHK} colors for main-sequence stars derived by \citet{jo66,jo68} and \citet{ko83}, respectively, are also applicable to the 2MASS {\it JHK}$_s$ system? Such a possibility can be tested from available 2MASS observations of standard stars and stars in open clusters of known reddening.

\begin{figure}[!t]
\begin{center}
\includegraphics[width=7.5cm]{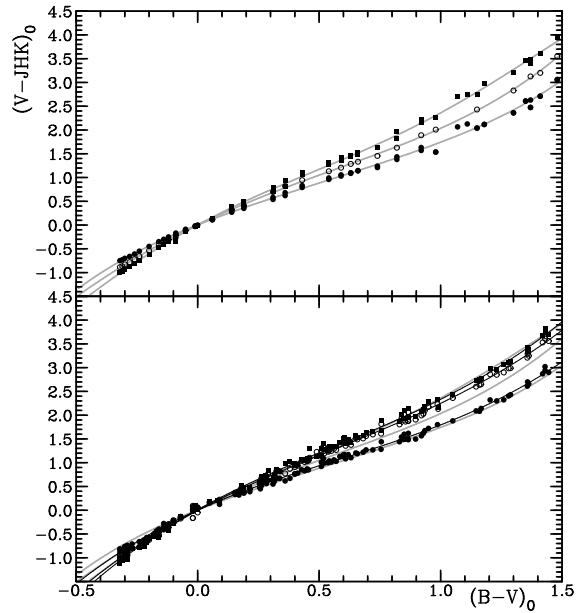}
\end{center}
\caption{\small{Intrinsic {\it (V--K)}$_0$ (filled squares), {\it (V--H)}$_0$ (open circles), and {\it (V--J)}$_0$ (filled circles) colors for main-sequence stars are plotted as functions of intrinsic {\it (B--V)}$_0$ color from \citet{jo68} and \citet{ko83} in the top portion of the diagram. The adopted relations from least squares fits are denoted by gray curves. The lower diagram plots observed data for photometric standards and standard stars using the same symbols as above. Solid curves denote best fits to the data, and differ from the predicted relations (gray curves).}}
\label{fig1}
\end{figure}

Intrinsic {\it JHK}$_s$ colors for main-sequence stars were initially derived here as follows. The relationships of \citet{jo66,jo68} and \citet{ko83} for intrinsic {\it V--J}, {\it V--H}, and {\it V--K} colors as functions of {\it (B--V)}$_0$ were tabulated and plotted, as in Fig.~\ref{fig1} (upper), including estimates for {\it V--H} color from the intrinsic colors of \citet{jo66,jo68} using his tabulated values for {\it V--J} and {\it V--K} interpolated according to the effective wavelengths of the filters. Least squares fits to the data then established polynomial relationships between the colors.

The derived relationships were then tested using 2MASS data and {\it BV} observations for 19 reasonably bright, nearby and unreddened, photometric and spectroscopic dwarf standards, stars with unsaturated observations in the Ursa Major and Hyades clusters, which are both unreddened, and stars in the young cluster NGC 2244 corrected for differential reddening within the Rosette Nebula \citep[see][]{tu76a}. NGC 2244 stars were used in order to tie down the hot end of the sequence; its member stars possess excellent photometry on the {\it UBV} system \citep{jo62} and lie in a region of well-established reddening law \citep{tu76a}. Reddening corrections for early-type stars were applied using {\it UBV} colors and the relations {\it E(J--H)} = 0.295 {\it E(B--V)}, {\it E(H--K}$_s$) = 0.49 {\it E(J--H)}, and $A_V = 2.427 E(J-H)$ for $R = A_V/E(B-V) = 3.05$, derived from van de Hulst's reddening curve No. 15 \citep[see][]{jo66}. In addition, the observations for Hyades stars were supplemented by data from \citet{ca82} in order to reduce the photometric scatter typical of 2MASS observations. The results are plotted in the lower portion of Fig.~\ref{fig1}.

The observational trends of {\it V--J}, {\it V--H}, and {\it V--K}$_s$ are fairly similar to the predicted trends seen in the top portion of Fig.~\ref{fig1}, but with noticeable offsets, particularly in {\it V-- H}. The adopted intrinsic relations were therefore established from polynomial fits to the observational data rather than from the predicted relations, with the results tabulated in Table~\ref{tab1} for zero-age main sequence (ZAMS) stars. The derived polynomial relationships of Fig.~\ref{fig1} were used to derive intrinsic {\it J--H} and {\it H--K} colors for main-sequence stars as a function of intrinsic broad band color index {\it (B--V)}$_0$, and the ZAMS calibration was that of \citet{tu76b,tu79}.

\begin{figure}[!t]
\begin{center}
\includegraphics[width=7cm]{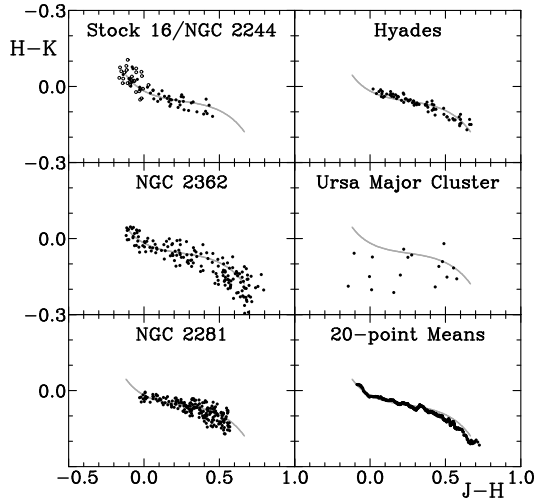}
\end{center}
\caption{\small{Unreddened {\it JHK}$_s$ colors (left) for stars in Stock 16 and NGC 2244 (top), NGC 2362 (middle), and NGC 2281 (bottom), and (right) for stars in the Hyades (top), and the Ursa Major cluster (middle) relative to the intrinsic relation for ZAMS stars (gray curves). The lower right diagram shows the data for stars in all clusters, except for saturated UMa stars, combined using running 20-point means.}}
\label{fig2}
\end{figure}

\setcounter{table}{0}
\begin{table*}
\caption[]{Empirical {\it JHK}$_s$ calibration for ZAMS stars.}
\label{tab1}
\centering
\small
\begin{tabular*}{0.94\textwidth}{@{\extracolsep{-1.4mm}}ccccccccccccccc}
\hline \noalign{\smallskip}
{\it J--H} &{\it H--K}$_s$ &$M_J$ &{\it J--H} &{\it H--K}$_s$ &$M_J$ &{\it J--H} &{\it H--K}$_s$ &$M_J$ &{\it J--H} &{\it H--K}$_s$ &$M_J$ &{\it J--H} &{\it H--K}$_s$ &$M_J$ \\
\noalign{\smallskip} \hline \noalign{\smallskip}
--0.129 &--0.107 &--2.393 &0.024 &0.023 &1.748 &0.148 &0.042 &3.035 &0.253 &0.058 &4.513 &0.391 &0.083 &5.438 \\
--0.125 &--0.104 &--2.021 &0.028 &0.026 &1.779 &0.151 &0.042 &3.088 &0.256 &0.058 &4.546 &0.396 &0.084 &5.455 \\
--0.121 &--0.100 &--1.749 &0.032 &0.027 &1.810 &0.154 &0.043 &3.142 &0.259 &0.059 &4.568 &0.400 &0.085 &5.481 \\
--0.117 &--0.097 &--1.506 &0.036 &0.027 &1.841 &0.157 &0.043 &3.196 &0.262 &0.059 &4.591 &0.405 &0.085 &5.507 \\
--0.113 &--0.094 &--1.303 &0.039 &0.027 &1.873 &0.160 &0.043 &3.239 &0.265 &0.060 &4.623 &0.409 &0.086 &5.533 \\
--0.109 &--0.090 &--1.130 &0.043 &0.028 &1.904 &0.162 &0.044 &3.293 &0.268 &0.061 &4.646 &0.414 &0.087 &5.559 \\
--0.105 &--0.087 &--0.976 &0.047 &0.028 &1.926 &0.165 &0.044 &3.337 &0.272 &0.061 &4.668 &0.419 &0.088 &5.584 \\
--0.101 &--0.083 &--0.832 &0.051 &0.029 &1.957 &0.168 &0.045 &3.390 &0.275 &0.062 &4.690 &0.424 &0.089 &5.609 \\
--0.097 &--0.080 &--0.678 &0.054 &0.029 &1.979 &0.171 &0.045 &3.434 &0.278 &0.062 &4.722 &0.428 &0.090 &5.633 \\
--0.093 &--0.076 &--0.533 &0.058 &0.030 &2.001 &0.173 &0.045 &3.478 &0.282 &0.063 &4.744 &0.433 &0.091 &5.658 \\
--0.088 &--0.072 &--0.388 &0.062 &0.030 &2.023 &0.176 &0.046 &3.521 &0.285 &0.063 &4.776 &0.438 &0.092 &5.681 \\
--0.084 &--0.069 &--0.233 &0.065 &0.031 &2.045 &0.179 &0.046 &3.565 &0.288 &0.064 &4.808 &0.443 &0.093 &5.705 \\
--0.080 &--0.065 &--0.088 &0.069 &0.031 &2.068 &0.182 &0.047 &3.609 &0.292 &0.065 &4.840 &0.448 &0.094 &5.728 \\
--0.119 &--0.045 &--2.279 &0.021 &0.024 &1.732 &0.188 &0.052 &2.982 &0.365 &0.069 &4.468 &0.536 &0.107 &5.370 \\
--0.116 &--0.042 &--1.912 &0.026 &0.025 &1.761 &0.193 &0.052 &3.035 &0.370 &0.070 &4.500 &0.540 &0.108 &5.386 \\
--0.112 &--0.040 &--1.644 &0.030 &0.026 &1.790 &0.198 &0.053 &3.089 &0.375 &0.070 &4.523 &0.544 &0.110 &5.411 \\
--0.109 &--0.037 &--1.405 &0.034 &0.028 &1.819 &0.202 &0.053 &3.143 &0.379 &0.071 &4.545 &0.549 &0.111 &5.437 \\
--0.106 &--0.035 &--1.206 &0.038 &0.029 &1.848 &0.207 &0.054 &3.186 &0.384 &0.072 &4.578 &0.553 &0.113 &5.462 \\
--0.102 &--0.032 &--1.037 &0.043 &0.030 &1.878 &0.212 &0.054 &3.240 &0.389 &0.072 &4.600 &0.557 &0.115 &5.486 \\
--0.099 &--0.030 &--0.887 &0.047 &0.031 &1.898 &0.217 &0.055 &3.284 &0.394 &0.073 &4.622 &0.562 &0.117 &5.511 \\
--0.095 &--0.027 &--0.748 &0.051 &0.032 &1.927 &0.221 &0.055 &3.338 &0.398 &0.074 &4.644 &0.566 &0.118 &5.535 \\
--0.092 &--0.025 &--0.597 &0.056 &0.033 &1.948 &0.226 &0.055 &3.382 &0.403 &0.074 &4.676 &0.570 &0.120 &5.558 \\
--0.088 &--0.023 &--0.457 &0.060 &0.034 &1.968 &0.231 &0.056 &3.426 &0.408 &0.075 &4.697 &0.574 &0.122 &5.582 \\
--0.085 &--0.020 &--0.316 &0.064 &0.034 &1.988 &0.236 &0.056 &3.470 &0.412 &0.076 &4.729 &0.579 &0.124 &5.605 \\
--0.081 &--0.018 &--0.164 &0.069 &0.035 &2.009 &0.240 &0.057 &3.514 &0.417 &0.077 &4.760 &0.583 &0.126 &5.628 \\
--0.077 &--0.016 &--0.023 &0.073 &0.036 &2.030 &0.245 &0.057 &3.558 &0.422 &0.078 &4.792 &0.587 &0.128 &5.650 \\
--0.074 &--0.014 &0.109 &0.078 &0.037 &2.061 &0.250 &0.058 &3.602 &0.427 &0.078 &4.823 &0.591 &0.130 &5.673 \\
--0.070 &--0.012 &0.231 &0.082 &0.038 &2.072 &0.255 &0.058 &3.646 &0.431 &0.079 &4.854 &0.595 &0.132 &5.694 \\
--0.066 &--0.010 &0.344 &0.087 &0.039 &2.093 &0.260 &0.058 &3.690 &0.436 &0.080 &4.884 &0.599 &0.135 &5.716 \\
--0.063 &--0.008 &0.447 &0.091 &0.039 &2.114 &0.264 &0.059 &3.724 &0.441 &0.081 &4.915 &0.603 &0.137 &5.737 \\
--0.059 &--0.006 &0.550 &0.096 &0.040 &2.146 &0.269 &0.059 &3.768 &0.445 &0.082 &4.945 &0.608 &0.139 &5.758 \\
--0.055 &--0.004 &0.654 &0.100 &0.041 &2.178 &0.274 &0.060 &3.802 &0.450 &0.083 &4.976 &0.612 &0.141 &5.779 \\
--0.051 &--0.002 &0.757 &0.105 &0.042 &2.200 &0.279 &0.060 &3.846 &0.454 &0.084 &5.006 &0.616 &0.144 &5.799 \\
--0.047 &--0.001 &0.852 &0.109 &0.042 &2.232 &0.283 &0.061 &3.880 &0.459 &0.085 &5.026 &0.620 &0.146 &5.819 \\
--0.044 &0.001 &0.946 &0.114 &0.043 &2.254 &0.288 &0.061 &3.923 &0.464 &0.086 &5.045 &0.624 &0.149 &5.838 \\
--0.040 &0.003 &1.041 &0.118 &0.044 &2.286 &0.293 &0.061 &3.957 &0.468 &0.087 &5.065 &0.628 &0.151 &5.857 \\
--0.036 &0.005 &1.116 &0.123 &0.044 &2.328 &0.298 &0.062 &3.991 &0.473 &0.088 &5.084 &0.631 &0.154 &5.876 \\
--0.032 &0.006 &1.201 &0.127 &0.045 &2.361 &0.303 &0.062 &4.035 &0.477 &0.089 &5.103 &0.635 &0.156 &5.895 \\
--0.028 &0.008 &1.266 &0.132 &0.046 &2.393 &0.308 &0.063 &4.069 &0.482 &0.090 &5.122 &0.639 &0.159 &5.913 \\
--0.024 &0.009 &1.342 &0.137 &0.046 &2.426 &0.312 &0.063 &4.102 &0.487 &0.091 &5.140 &0.643 &0.162 &5.930 \\
--0.020 &0.011 &1.408 &0.141 &0.047 &2.469 &0.317 &0.064 &4.136 &0.491 &0.093 &5.159 &0.647 &0.164 &5.948 \\
--0.016 &0.012 &1.465 &0.146 &0.047 &2.502 &0.322 &0.064 &4.169 &0.496 &0.094 &5.177 &0.651 &0.167 &5.965 \\
--0.012 &0.014 &1.521 &0.151 &0.048 &2.555 &0.327 &0.065 &4.203 &0.500 &0.095 &5.195 &0.655 &0.170 &5.981 \\
--0.008 &0.015 &1.558 &0.155 &0.048 &2.598 &0.332 &0.065 &4.236 &0.505 &0.096 &5.213 &0.658 &0.173 &5.997 \\
--0.004 &0.017 &1.585 &0.160 &0.049 &2.651 &0.336 &0.066 &4.270 &0.509 &0.098 &5.240 &0.662 &0.176 &6.013 \\
0.000 &0.018 &1.612 &0.165 &0.049 &2.704 &0.341 &0.066 &4.303 &0.514 &0.099 &5.258 &0.666 &0.179 &6.028 \\
0.005 &0.019 &1.640 &0.169 &0.050 &2.748 &0.346 &0.067 &4.336 &0.518 &0.101 &5.285 & $\cdots$ & $\cdots$ & $\cdots$ \\
0.009 &0.021 &1.658 &0.174 &0.050 &2.811 &0.351 &0.067 &4.379 &0.523 &0.102 &5.301 & $\cdots$ & $\cdots$ & $\cdots$ \\
0.013 &0.022 &1.676 &0.179 &0.051 &2.874 &0.355 &0.068 &4.412 &0.527 &0.103 &5.328 & $\cdots$ & $\cdots$ & $\cdots$ \\
0.017 &0.023 &1.704 &0.183 &0.051 &2.928 &0.360 &0.069 &4.445 &0.531 &0.105 &5.344 & $\cdots$ & $\cdots$ & $\cdots$ \\
\noalign{\smallskip} \hline
\end{tabular*}
\end{table*}

The intrinsic relationship for 2MASS colors in Table~\ref{tab1} is similar to the intrinsic color-color relation for dwarf stars in the {\it UBV} system, displaying a noticeable ``kink'' near spectral type A0. 2MASS {\it J} magnitudes sample the Brackett continuum in hot stars, while {\it H} and {\it K}$_s$ magnitudes sample the Pfund continuum, so {\it J--H} color should provide a measure of the Pfund discontinuity in hot stars, much like {\it U--B} color provides a measure of the Balmer discontinuity. Similarly, {\it H--K}$_s$ color should provides a measure of stellar temperature, much like {\it B--V} color does in the {\it UBV} system. For cool stars all colors should closely track changes in the slope of the black body continuum in the far infrared. Interstellar reddening affects {\it JHK}$_s$ colors much less than it does {\it UBV} colors, but the effects of circumstellar emission are often more important in the far infrared than in the visible region. A {UBV} color-color diagram plots {\it U--B} versus {\it B--V}, for which a plot of {\it J--H} versus {\it H--K}$_s$ would be the closest approximation. It appears, however, that {\it J--H} colors may display a greater precision than {\it H--K}$_s$ colors in 2MASS photometry, so the diagnostic tool used here to establish reddening for cluster stars is a 2MASS color-color diagram in which {\it H--K}$_s$ is plotted versus {\it J--H}.

The intrinsic relations were tested further using the three clusters used for the calibration and three additional clusters that lie in fields of uniform interstellar reddening: Stock 16 \citep{tu85a}, NGC 2362 \citep{jo53,jo57,jo61}, and NGC 2281 \citep{pe61}, which are reddened by {\it E(B--V)} = 0.49, 0.11, and 0.11, respectively \citep[see][]{tu96,jo61}. Obvious unreddened stars in the same fields were also included in the samples. Fig.~\ref{fig2} plots 2MASS colors for stars in the core regions of Stock 16, NGC 2362, and NGC 2281 with cited magnitude uncertainties of less than $\pm0.05$, corrected for their known reddening or lack of reddening for suspected foreground stars, along with similar data for Hyades and Ursa Major cluster stars and dereddened data for stars in NGC 2244. Data for the Ursa Major cluster include 11 stars with observations of uncertain quality (bright stars with saturated photometry and blended stars). The bottom right portion of Fig.~\ref{fig2} displays the data for all clusters, except for UMa stars of poor quality, combined using running 20-point means as a function of {\it J--H} color. The last step was made as a means of reducing the photometric scatter in both colors.

The intrinsic relation of Table~\ref{tab1} appears to be confirmed by stars in the open clusters of Fig.~\ref{fig2}, although there is clearly additional scatter in the observations of individual stars that cannot be attributed to differential reddening. In three of the four reddened clusters the {\it UBV} colors display {\it no} differential reddening, so it is unlikely to appear in the {\it JHK}$_s$ colors unless there is some unforeseen component, such as circumstellar emission, affecting the observations. A pronounced ``extra'' scatter for B-type stars may have such an origin, given the evidence for circumstellar dust associated with some rapidly-rotating late B-type stars \citep{tu93}, but it is unlikely that similar effects extend to the remaining cluster stars. It is more reasonable to attribute the scatter to intrinsic uncertainties in the precision of the {\it JHK}$_s$ magnitude estimates, which appear to be larger than the cited values and also larger than what is attainable from iris photometry of photographic plates \citep{tw89}. The tendency toward larger scatter for the coolest stars of NGC 2362 and NGC 2281 may be a problem with membership selection.

An earlier calibration similar to that of Table~\ref{tab1} was derived using stars in Stock 16, NGC 2362, and NGC 2251, with polynomial and linear fits made to combined running 20-point means for the data. Some of the features of the earlier calibration \citep[see][]{te08,te09} can be seen in the lower right portion of Fig.~\ref{fig2}, namely the pronouned ``kink'' near spectral type A0 in the mean data. But the data also match the new curved relationship very well.

For Hyades stars the {\it JHK}$_s$ observations represent a combination of both 2MASS data and data from \citet{ca82}, as noted. The colors from both data sets display a similar scatter to that for Stock 16 and NGC 2244 members, but is much reduced when the data are averaged together. There remains a sizeable residual scatter that is slightly larger than the cited uncertainties in the observations. Perhaps it represents an observational limitation for ground-based observations in the infrared imposed by variable atmospheric water vapour content \citep[e.g.][]{my05,my08}. The scatter is greatly reduced in the 20-point running means, which is consistent with a problem that is tied to the precision of the observations. Otherwise, the intrinsic relation of Table~\ref{tab1} appears to be confirmed.

The colors for cluster stars in Fig.~\ref{fig2} agree closely with the derived intrinsic relation, in most cases displaying only random scatter about it. If there were an extra reddening component of, say, 0.01 to 0.02 in {\it E(J--H)} for the stars, the centroids for the data would display a noticeable offset from the intrinsic relation. Thus, despite large scatter in the colors for stars in many open clusters, it is possible to derive an observational reddening for the stars from a color-color diagram by eye, with uncertainties of no more than $\pm 0.01$ to $\pm 0.02$ in {\it E(J--H)}, provided that the scatter in such cases is random in nature. Likewise, distance moduli derived from observational {\it J} versus {\it J--H} color-magnitude diagrams can be derived by eye with uncertainties typically no larger than $\pm 0.1$ to $\pm 0.2$. Such conclusions are confirmed by tests on the many open clusters studied to date \citep[see, for example,][]{te08,te09}.
 
\section{Other Open Clusters}
As noted above, the intrinsic relations of Table~\ref{tab1} have been tested successfully previously \citep{te08,te09}, but three clusters of uncertain properties were selected for further testing here: Berkeley 44, Turner 1, and Collinder 419. All three clusters possess limited published optical observations, and provide good examples of where 2MASS observations may help to clarify previous conclusions about cluster properties.

\begin{figure}[!t]
\begin{center}
\includegraphics[width=6.2cm]{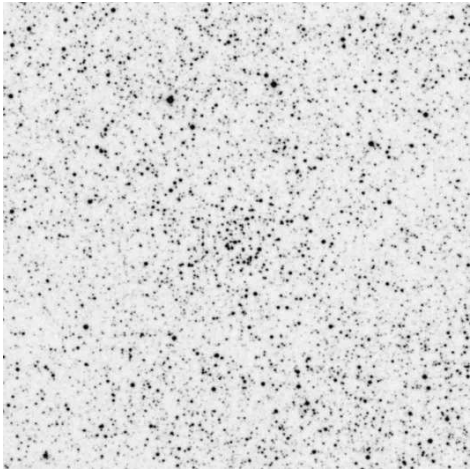}
\end{center}
\caption{\small{The $15\arcmin \times 15\arcmin$ field of Berkeley 44 from the Palomar Observatory Sky Survey red image of the field. The image is centered on 2000 co-ordinates: 19:17:16, +19:33:00, the cluster center identified visually.}}
\label{fig3}
\end{figure}

\begin{figure}[!t]
\begin{center}
\includegraphics[width=6.2cm]{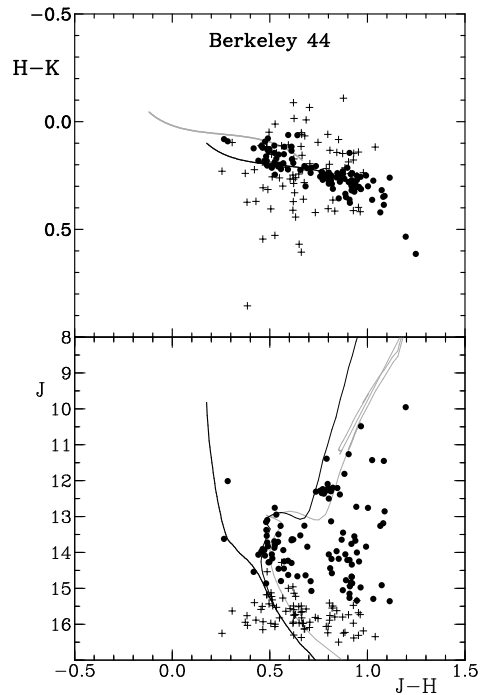}
\end{center}
\caption{\small{{\it JHK}$_s$ photometry for stars within $2\arcmin$ of the center of Berkeley 44 with magnitude uncertainties smaller than $\pm0.05$ (points) as well as with larger uncertainties (plus signs). The intrinsic color-color relation is depicted in gray (top), while the intrinsic relations for {\it E(J--H)} = 0.295 ({\it E(B--V)} = 1.00) and {\it J--M}$_J$ = 12.1 ({\it V}$_0${\it--M}$_V$ = 11.38) are shown as black lines. The thin black curve is an isochrone for $\log t = 9.3$ adapted from \citet{me93}, the thin gray curve an isochrone of identical age from \citet{bo04}.}}
\label{fig4}
\end{figure}

\subsection{Berkeley 44}
The sparse northern hemisphere cluster Berkeley 44 (Fig.~\ref{fig3}) has been identified as an old group by \citet*{ca06}, although with some question about its reality because of ambiguities in the star counts and the similarity of its color-magnitude diagram to that of stars in the surrounding reference field. The published co-ordinates for the cluster by \citet{di02} and \citet{ca06} do not match the optical density peak visible on the Palomar Observatory Sky Survey, and an alternate cluster center (Fig.~\ref{fig3}) was adopted here. The group lies more than $3^{\circ}$ from the Galactic plane, so contamination by foreground early-type stars should be relatively low. However, the extreme faintness of cluster stars makes it necessary to consider both high quality and low quality 2MASS data for the field, with uncertainties of $\pm0^{\rm m}.05$ representing the demarcation.

The 2MASS {\it JHK}$_s$ data for Berkeley 44 (Fig.~\ref{fig4}) confirm that it is an old open cluster. The cluster color-color diagram is devoid of early-type stars lying within $2\arcmin$ of the  adopted cluster center, most cluster members being cooler late-type stars. A few stars of inferred spectral types G and K appear to be essentially unreddened, with most stars of spectral types F or later reddened by similar amounts. The cluster color-magnitude diagram reveals a well-defined clump of red giants at $J \simeq 12.2$ with implied spectral types of early K, so the identification of this cluster as an old cluster is confirmed. The optimum fit by eye to the {\it JHK}$_s$ observations yields a reddening of {\it E(J--H)} $=0.295\pm0.02$ ({\it E(B--V)} $ = 1.00 \pm0.07$) and a distance modulus of {\it J--M}$_J = 12.1 \pm0.1$ ({\it V}$_0${\it--M}$_V = 11.38 \pm0.24$), corresponding to a distance of $d = 1.89 \pm0.21$ kpc. The uncertainty in the intrinsic distance modulus includes the uncertainty in interstellar extinction towards the cluster. The ZAMS fit is tied to low quality {\it JHK}$_s$ observations at the faint end, so may be less well-established than the formal uncertainty suggests, although the implied distance agrees closely with the value of 1.8 kpc derived by \citet{ca06}. The implied cluster reddening is significantly smaller than the value of {\it E(B--V)} = 1.40 $\pm0.10$ obtained by \citet{ca06}, which in any case is inconsistent with the 2MASS data unless one adopts an extreme fit to the very reddest stars.

Berkeley 44 main-sequence members have intrinsic {\it J--H} colors of $\sim0.15$ or redder, indicating stars of earliest spectral type $\sim$F0, corresponding to turnoffs for intermediate-age clusters of age $\sim10^9$ yrs. A variety of model isochrones near $\log t \simeq 9$ were therefore tested in the cluster color-magnitude diagram (Fig.~\ref{fig4}), with the optimum fit produced by an isochrone of $\log t = 9.3$, as shown, corresponding to a cluster age of $\sim2$ Gyr. The procedure involved use of the {\it M}$_V$ versus {\it B--V} isochrones published by \citet*{me93}, adapted to the {\it M}$_J$ versus {\it J--H} system with our empirical link between the {\it UBV} and 2MASS systems. An isochrone of the same age from \citet*{bo04} is also plotted in Fig.~\ref{fig4}, but deviates systematically in {\it J--H} from the \citet{me93} isochrone, suggesting a possible problem in the conversion from $\log T_{\rm eff}$ to {\it J--H} with the Padova isochrones used by \citet{bo04}. A problem arises with stars in the red giant clump, since they are not replicated well by either set of evolutionary models used for the isochrones. Yet their colors are consistent with the adopted cluster reddening. Berkeley 44 is indeed an old open cluster, but inferences about its parameters derived using the larger reddening estimated by \citet{ca06} differ from those inferred from 2MASS observations.

\begin{figure}[!t]
\begin{center}
\includegraphics[width=6.2cm]{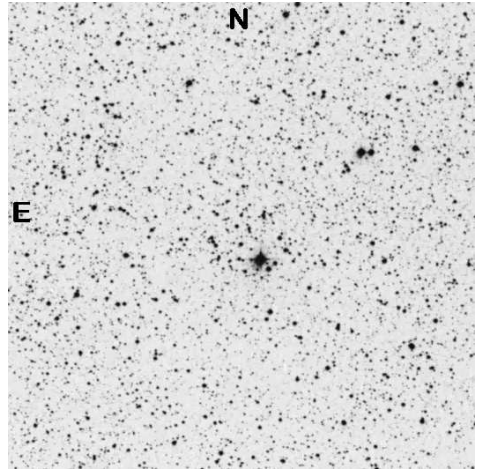}
\end{center}
\caption{\small{The $15\arcmin \times 15\arcmin$ field of Turner 1 from the Palomar Observatory Sky Survey red image of the field. The image is centered on 2000 co-ordinates: 19:48:26.5, +27:17:59, the cluster center identified from star counts \citep{tu85b}. S Vul is the bright star near the center of the image.}}
\label{fig5}
\end{figure}

\begin{figure}[!t]
\begin{center}
\includegraphics[width=6.2cm]{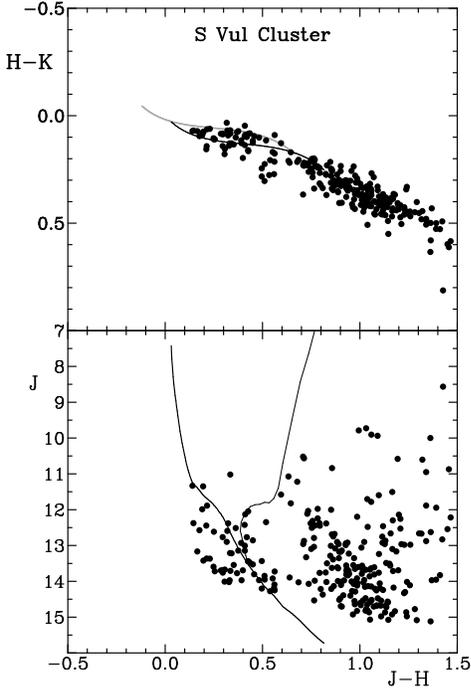}
\end{center}
\caption{\small{{\it JHK}$_s$ photometry for stars within $4\arcmin$ of the center of Turner 1 with magnitude uncertainties smaller than $\pm0.05$. The intrinsic color-color relation is depicted in gray, while the intrinsic relations for {\it E(J--H)} = 0.15 ({\it E(B--V)} = 0.51) and {\it J--M}$_J$ = 9.7 ({\it V}$_0${\it--M}$_V$ = 9.34) are shown as black lines. The thin black curve is an isochrone for $\log t = 9.8$ adapted from \citet{me93}.}}
\label{fig6}
\end{figure}

\begin{figure}[!t]
\begin{center}
\includegraphics[width=6.2cm]{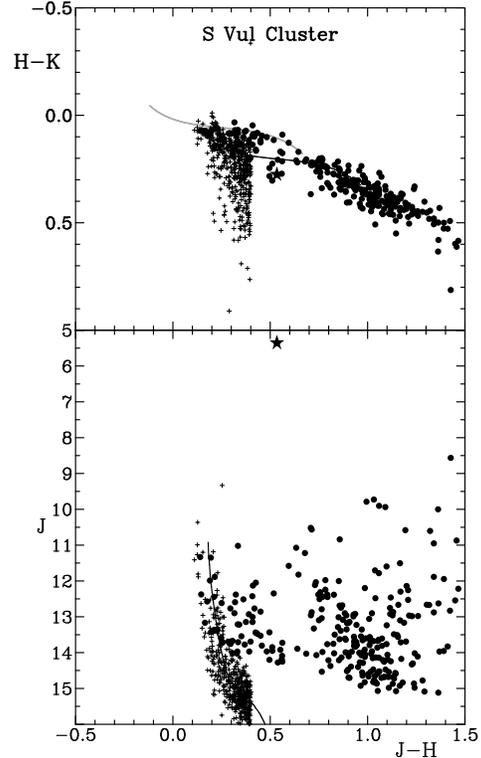}
\end{center}
\caption{\small{The same as Fig.~\ref{fig6} but for {\it E(J--H)} = 0.29 ({\it E(B--V)} = 1.02) and {\it J--M}$_J$ = 13.2 ({\it V}$_0${\it--M}$_V$ = 12.47), shown as black lines. Small plus signs denote additional stars within $11\arcmin$ of the cluster center inferred to be B-type and A-type stars, with no restriction on photometric uncertainties. The star symbol corresponds to the Cepheid S Vul.}}
\label{fig7}
\end{figure}

\subsection{Turner 1}
The sparse cluster designated as Turner 1 (Fig.~\ref{fig5}) surrounding the long-period Cepheid S Vul was discovered during the preparation of finder charts for Galactic Cepheids \citep{tu85b}, and was subsequently studied on the basis of photoelectric and photographic {\it UBV} photometry by \citet*{te86}.

{\it UBV} photometry for stars in Turner 1 yielded a cluster distance of $643 \pm25$ pc and a reddening of {\it E(B--V)} $= 0.48 \pm0.02$ for what appeared to be an old group of G-type dwarfs with a putative red giant branch, although there was also evidence for more heavily reddened B-type stars in the same field  \citep{te86}. The limited {\it UBV} data implied that Turner 1 lay foreground to the Cepheid S Vul, which is projected near its center, although no further tests were made, spectroscopic observations being restricted by the faintness of cluster stars. It was noted, however, that the putative reddened G-dwarf sequence identified from {\it UBV} colors might be contaminated by more heavily reddened B dwarfs.

2MASS {\it JHK}$_s$ data provide new insights into the earlier results. It is possible to detect a sparse group of G dwarfs (with rather large photometric scatter) and a putative late-K giant branch in the data, with similar parameters to the {\it UBV} study (Fig.~\ref{fig6}). But many of the stars in the field, and most of the stars within $11\arcmin$ of the adopted cluster center, appear instead to be reddened B-type and A-type stars (Fig.~\ref{fig7}). A best fit by eye to the {\it JHK}$_s$ observations for the former yields a reddening of {\it E(J--H)} $= 0.15 \pm0.02$ ({\it E(B--V)} $= 0.51 \pm0.07$) and a distance modulus of {\it J--M}$_J = 9.7 \pm0.1$ ({\it V}$_0${\it--M}$_V = 9.34 \pm0.24$), corresponding to a distance of $d = 0.74 \pm0.08$ kpc, values that are reasonably consistent with those obtained for the cluster from {\it UBV} photometry. The model isochrone from \citet{me93} that best fits the observations has $\log t = 9.8$, implying a cluster age of $\sim6$ Gyr, although the red giants are displaced redward of that isochrone, much like the case for Berkeley 44 red giant clump stars. Conceivably a refined empirical calibration of 2MASS intrinsic colors that includes a separate relation for red giant stars could eliminate the problem. There appear to be large numbers of M dwarfs lying along the line of sight to the cluster.

The implied best fit by eye to the {\it JHK}$_s$ observations for the group of reddened B-type stars yields a reddening of {\it E(J--H)} $= 0.30 \pm0.02$ ({\it E(B--V)} $= 1.02 \pm0.07$) and a distance modulus of {\it J--M}$_J = 13.2 \pm0.2$ ({\it V}$_0${\it--M}$_V = 12.47 \pm0.29$), corresponding to a distance of $d = 3.12 \pm0.42$ kpc. Similar values were used by \citet{tu10} with an earlier version of the 2MASS calibration to establish very reasonable estimates for the reddening and luminosity of the 68$^{\rm d}$ Cepheid S Vulpeculae as a possible cluster member.

S Vul is represented by the star symbol in Fig.~\ref{fig7}. In the present instance the results imply intrinsic parameters of $(\langle B \rangle - \langle V \rangle)_0 = 0.87 \pm 0.07$ and $\langle M_V \rangle = -6.86 \pm 0.29$ for the Cepheid, close to what would be predicted empirically \citep[see][]{tu10}. The angular brackets denote intensity means. Spectroscopic observations are essential for confirming the picture implied by the 2MASS observations, since star counts for stars identified photometrically as B dwarfs display only a marginal concentration towards the cluster core. It appears that the main concentration of stars in Turner 1 corresponds to the old, sparse, foreground cluster identified in Fig.~\ref{fig6}, with contamination by a young group of background B-dwarfs located along the same line of sight.

\begin{figure}[!t]
\begin{center}
\includegraphics[width=6.2cm]{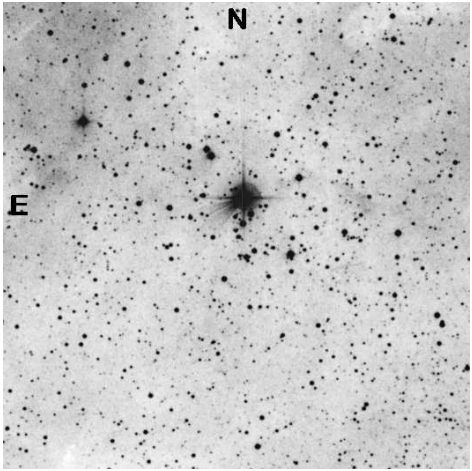}
\end{center}
\caption{\small{The $15\arcmin \times 15\arcmin$ field of Collinder 419 from the Palomar Observatory Sky Survey red image of the field. The image is centered on 2000 co-ordinates: 20:18:08, +40:42:42, the adopted cluster center. HD 193322 is the bright star north of the center of the image.}}
\label{fig8}
\end{figure}

\begin{figure}[!t]
\begin{center}
\includegraphics[width=6.2cm]{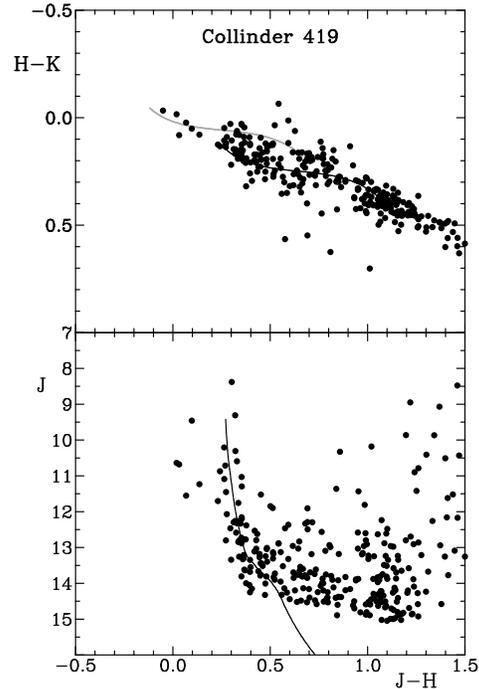}
\end{center}
\caption{\small{{\it JHK}$_s$ photometry for stars within $5\arcmin$ of the adopted center for Collinder 419 with uncertainties smaller than $\pm0.05$. The intrinsic color-color relation is depicted in gray, while the intrinsic relations for {\it E(J--H)} = 0.39 ({\it E(B--V)} = 1.32) and {\it J--M}$_J$ = 11.7 ({\it V}$_0${\it--M}$_V$ = 10.75) are shown as black lines.}}
\label{fig9}
\end{figure}

\subsection{Collinder 419}
Collinder 419 is a sparse group of stars (Fig.~\ref{fig8}) of small angular diameter \citep[$4\arcmin$.5 according to][]{co31} associated with the massive O9 V((n)) double star HD 193322 \citep{wa71}. Collinder estimated a distance of 1470 pc for the group from its angular diameter, while \citet{gi10} used astrometric and photometric data from the UCAC3 catalogue for bright stars in the field to derive a distance of $d = 741 \pm36$ pc, with a reddening of {\it E(B--V)} $= 0.37 \pm0.05$. Included as a possible member of the group was the M3 III star IRAS 20161+4035.

\setcounter{table}{1}
\begin{table*}
\caption[]{Deduced parameters for clusters analyzed by {\it JHK}$_s$ photometry.}
\label{tab2}
\centering
\small
\begin{tabular*}{0.75\textwidth}{@{\extracolsep{+0.8mm}}lccccc}
\hline \noalign{\smallskip}
Cluster &{\it E(J--H)} &{\it E(B--V)} &{\it J--M}$_J$ &{\it V}$_0${\it--M}$_V$ &{\it d} (kpc) \\
\noalign{\smallskip} \hline \noalign{\smallskip}
Berkeley 44 &$0.295 \pm0.02$ &$1.00 \pm0.07$ &$12.1 \pm0.1$ &$11.38 \pm0.24$ &$1.89 \pm0.21$ \\
Turner 1a &$0.15 \pm0.02$ &$0.51 \pm0.07$ &$9.7 \pm0.2$ &$9.34 \pm0.29$ &$0.74 \pm0.08$ \\
Turner 1b &$0.29 \pm0.02$ &$1.02 \pm0.07$ &$13.2 \pm0.2$ &$12.47 \pm0.29$ &$3.12 \pm0.42$ \\
Collinder 419 &$0.39 \pm0.02$ &$1.32 \pm0.07$ &$11.7 \pm0.2$ &$10.75 \pm0.28$ &$1.42 \pm0.18$ \\
\noalign{\smallskip} \hline
\end{tabular*}
\end{table*}

The cluster does not stand out well at optical wavelengths, but we were able to detect a slight density enhancement visually and estimate a crude center of symmetry at 2000 co-ordinates: 20:18:08, +40:42:42, close to the coordinates estimated by Collinder. The cluster also appears as a set of reddened B-type stars in available 2MASS {\it JHK}$_s$ photometry for stars lying within $5\arcmin$ of that center of symmetry (Fig.~\ref{fig9}), reasonably consistent with Collinder's inferred dimensions. A best fit by eye to the {\it JHK}$_s$ data yields a reddening of {\it E(J--H)} $= 0.39 \pm0.02$ ({\it E(B--V)} $= 1.32 \pm0.07$), significantly larger than the value estimated by \citet{gi10}, a distance modulus of {\it J--M}$_J = 11.7 \pm0.2$ ({\it V}$_0${\it--M}$_V = 10.75 \pm0.28$), and an inferred distance of $d = 1.42 \pm0.18$ kpc, which is also larger than the \citet{gi10} value, although consistent with a location within the young complex of stars and H II regions \citep[including Berkeley 87,][]{te10} associated with the Cygnus X region at $\sim1.2$ kpc. In this case the 2MASS data provide a more solid basis for the reality of the group than the UCAC3 data, although they do not address the possible membership of HD 193322.

The cluster does not show up as a significant density enhacement from star counts, and it is conceivable that Collinder 419 represents merely a clump in one of the many OB associations seen along the direction of the Cygnus arm.

\section{Discussion}
Table~\ref{tab2} summarizes the results of the present study of four open clusters using 2MASS observations. Most of the inferred parameters for the clusters differ from published results tied to optical photometry, although those for Turner 1 were considered as an alternate possibility in the study by \citet{te86}. In each case the optical photometry was limited by the faintness of cluster members arising from large interstellar reddening, which is where {\it JHK}$_s$ photometry has advantages over optical band photometry. The point to emphasize, however, is that {\it JHK}$_s$ observations of cluster stars often provide relatively complete samples that can be used to infer reasonably accurate estimates for the all-important interstellar reddening toward the cluster, in most cases in more straightforward manner than what is often used for optical band {\it BVRI} observations. Once the reddening is known, ZAMS fitting, or isochrone fitting in some cases, can then be used to derive the distance to the cluster. That is true despite the relatively low intrinsic precision of existing 2MASS data for most Galactic fields.

\section*{acknowledgments}
\small{This publication makes use of data products from the Two Micron All Sky Survey, which is a joint project of the University of Massachusetts and the Infrared Processing and Analysis Center/California Institute of Technology, funded by the National Aeronautics and Space Administration and the National Science Foundation.}

\end{document}